\documentclass[twoside,twocolumn,9pt]{article}
\usepackage{extsizes}
\usepackage{multirow}
\usepackage[super,sort&compress,comma]{natbib} 
\usepackage[version=3]{mhchem}
\usepackage[left=1.5cm, right=1.5cm, top=1.785cm, bottom=2.0cm]{geometry}
\usepackage{balance}
\usepackage{mathptmx}
\usepackage{amsmath}
\usepackage{sectsty}
\usepackage{graphicx} 
\usepackage{lastpage}
\usepackage[format=plain,justification=justified,singlelinecheck=false,font={stretch=1.125,small,sf},labelfont=bf,labelsep=space]{caption}
\usepackage{float}
\usepackage{fancyhdr}
\usepackage{fnpos}
\usepackage[english]{babel}
\addto{\captionsenglish}{%
  
}
\usepackage{array}
\usepackage{droidsans}
\usepackage{charter}
\usepackage[T1]{fontenc}
\usepackage[usenames,dvipsnames]{xcolor}
\usepackage{setspace}
\usepackage[compact]{titlesec}
\usepackage{hyperref}
\usepackage{amssymb}

\newcommand{\cbb}{\color{black}}

\newcommand{\rr}{{\bf r}}

\newcommand{\BE}{\begin{equation}}
\newcommand{\EE}{\end{equation}}

\usepackage{epstopdf}

\definecolor{cream}{RGB}{222,217,201}

\begin{document}

\pagestyle{fancy}
\thispagestyle{plain}
\fancypagestyle{plain}{
\renewcommand{\headrulewidth}{0pt}
}

\makeFNbottom
\makeatletter
\renewcommand\LARGE{\@setfontsize\LARGE{15pt}{17}}
\renewcommand\Large{\@setfontsize\Large{12pt}{14}}
\renewcommand\large{\@setfontsize\large{10pt}{12}}
\renewcommand\footnotesize{\@setfontsize\footnotesize{7pt}{10}}
\makeatother

\renewcommand{\thefootnote}{\fnsymbol{footnote}}
\renewcommand\footnoterule{\vspace*{1pt}%
\color{cream}\hrule width 3.5in height 0.4pt \color{black}\vspace*{5pt}} 
\setcounter{secnumdepth}{5}

\makeatletter 
\renewcommand\@biblabel[1]{#1}            
\renewcommand\@makefntext[1]%
{\noindent\makebox[0pt][r]{\@thefnmark\,}#1}
\makeatother 
\renewcommand{\figurename}{\small{Fig.}~}
\sectionfont{\sffamily\Large}
\subsectionfont{\normalsize}
\subsubsectionfont{\bf}
\setstretch{1.125} 
\setlength{\skip\footins}{0.8cm}
\setlength{\footnotesep}{0.25cm}
\setlength{\jot}{10pt}
\titlespacing*{\section}{0pt}{4pt}{4pt}
\titlespacing*{\subsection}{0pt}{15pt}{1pt}

\fancyfoot{}
\fancyfoot[LO,RE]{\vspace{-7.1pt}\includegraphics[height=9pt]{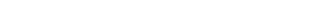}}
\fancyfoot[CO]{\vspace{-7.1pt}\hspace{13.2cm}\includegraphics{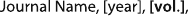}}
\fancyfoot[CE]{\vspace{-7.2pt}\hspace{-14.2cm}\includegraphics{head_foot/RF}}
\fancyfoot[RO]{\footnotesize{\sffamily{1--\pageref{LastPage} ~\textbar  \hspace{2pt}\thepage}}}
\fancyfoot[LE]{\footnotesize{\sffamily{\thepage~\textbar\hspace{3.45cm} 1--\pageref{LastPage}}}}
\fancyhead{}
\renewcommand{\headrulewidth}{0pt} 
\renewcommand{\footrulewidth}{0pt}
\setlength{\arrayrulewidth}{1pt}
\setlength{\columnsep}{6.5mm}
\setlength\bibsep{1pt}

\makeatletter 
\newlength{\figrulesep} 
\setlength{\figrulesep}{0.5\textfloatsep} 

\newcommand{\topfigrule}{\vspace*{-1pt}%
\noindent{\color{cream}\rule[-\figrulesep]{\columnwidth}{1.5pt}} }

\newcommand{\botfigrule}{\vspace*{-2pt}%
\noindent{\color{cream}\rule[\figrulesep]{\columnwidth}{1.5pt}} }

\newcommand{\dblfigrule}{\vspace*{-1pt}%
\noindent{\color{cream}\rule[-\figrulesep]{\textwidth}{1.5pt}} }

\makeatother

\twocolumn[ 
  \begin{@twocolumnfalse}
{\includegraphics[height=30pt]{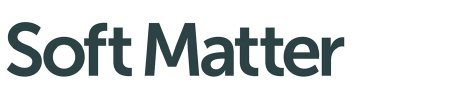}\hfill\raisebox{0pt}[0pt][0pt]{\includegraphics[height=55pt]{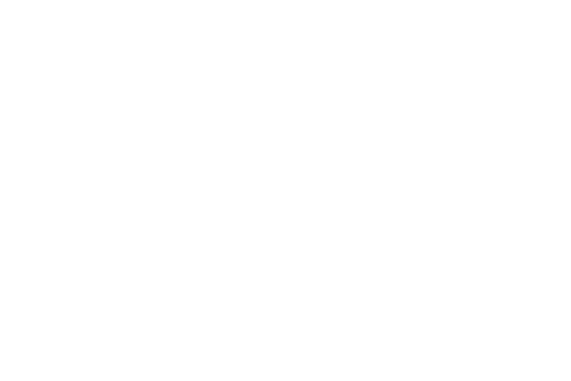}}\\[1ex]
\includegraphics[width=18.5cm]{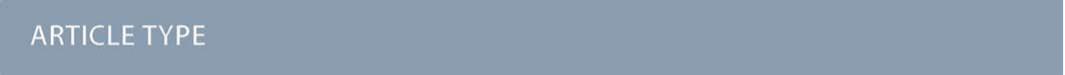}}\par
\vspace{1em}
\sffamily
\begin{tabular}{m{4.5cm} p{13.5cm} }

\includegraphics{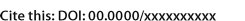} & \noindent\LARGE{\textbf{Stress--Boundary--Memory Feedback Drives Vortical--Polar Transitions in Softly Confined Active Matter$^\dag$}} \\
\vspace{0.3cm} & \vspace{0.3cm} \\

 & \noindent\large{Haosheng Wen,\textit{$^{a,b,c}$} P.B. Sunil Kumar,\textit{$^{d}$} and Mohamed Laradji\textit{$^{c\ast}$}}
 \\

\includegraphics{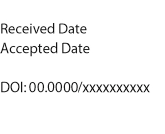} & \noindent\normalsize{
We computationally investigate how environmental sensitivity of active matter interacts with soft confinement to shape collective dynamics.
In our model, the active constituents are represented as self--propelled particles (SPPs), implemented as nematic, disjoint ring polymers
whose direction of motion can reverse without tumbling, with a directional persistence controlled by the driving force, ${F}_D$, and a persistence time scale, $\tau_m$. 
Coarse--grained molecular dynamics simulations of these reversal--capable SPPs confined within a deformable two--dimensional enclosure reveal that the collective dynamics arise from a three--way feedback between active stresses, boundary elasticity, and 
particle--level memory.
With increasing $F_D$, this stress–boundary–memory feedback generates a sequence of collective dynamical regimes. At low ${F}_D$, SPP motion is dominated by thermal fluctuations and activity plays a negligible role. At intermediate ${F}_D$, coherent vortical motion emerges with intermittent, noise--driven reversals. The frequency of reversals is modulated by boundary elasticity and $\tau_m$, and their occurrence coincides with transient coherent polar motion. 
With further increase in ${F}_D$, reversals are suppressed, yielding sustained unidirectional vortical motion in which the enclosure exhibits diffusive propulsion with a diffusivity that varies non--monotonically with ${F}_D$.
At sufficiently high $F_D$, the system transitions to a polar state characterized by strong nematic ordering of the SPPs, symmetry breaking of the enclosure shape, and persistent polar collective motion. In this regime, the SPPs accumulate at the leading edge of the enclosure, deforming it into an anisotropic shape and driving sustained ballistic propulsion of the enclosure with a slowly drifting direction. These results demonstrate how environmental sensitivity and soft confinement jointly regulate emergent collective states of confined active matter and identify boundary elasticity as a control parameter governing the balance between vortical and ballistic dynamics.}

\\

\end{tabular}

 \end{@twocolumnfalse} \vspace{0.6cm}


\renewcommand*\rmdefault{bch}\normalfont\upshape
\rmfamily
\section*{}
\vspace{-1cm}]

\footnotetext{\textit{$^{a}$~Department of Biochemistry and Molecular Biology, University of Chicago, Chicago, IL 60637, USA}}
\footnotetext{\textit{$^{b}$~Center for Mechanical Excitability, University of Chicago, Chicago, IL 60637, USA}}
\footnotetext{\textit{$^{c}$~Department of Physics and Materials Science, The University of Memphis, Memphis, TN 38152, USA}}
\footnotetext{\textit{$^{d}$~Department of Physics, Indian Institute of Technology Madras, Chennai-600036, Tamil Nadu, India}}

\footnotetext{\textit{$^\ast$~mlaradji@memphis.edu}}

\footnotetext{\dag~Electronic Supplementary Information (ESI) available  See DOI: 10.1039/cXsm00000x/}




\section{Introduction}

Active matter systems, composed of energy--consuming units that generate persistent motion, 
exhibit a wide range of nonequilibrium collective behaviors, including  swarming,  flocking, coherent rotations, oscillations, clustering, and motility--induced phase separation~\cite{schweitzer07,Marchetti2013,Sriram2017,Bechinger2016}.
Such units, commonly referred to as self--propelled particles (SPPs), are found in many natural and engineered systems. Examples include
actin filaments and microtubules
~\cite{vicsek1995novel, szabo2006phase, wang2011spontaneous, mehes2014collective}, dense colonies of bacteria and eukaryotic cells~\cite{vicsek1995novel, szabo2006phase, wang2011spontaneous, mehes2014collective},
active colloids~\cite{theurkauff12}, and larger--scale systems such as schools of fish~\cite{scott2001}, flocks of birds~\cite{bialek12}, human crowds~\cite{corbetta23}, granular assemblies~\cite{blair2003vortices,kudrolli2008swarming,lam2015self}, deformable cell--mimicking structures containing granular particles~\cite{arora2024}, and robots following simple rules~\cite{deblais2018boundaries, boudet2021collections}. Despite differences in microscopic details, these systems often exhibit recurring patterns of collective organization.

The collective dynamics of SPPs depends on their motility, density, and alignment interactions~\cite{Marchetti2013, Cates2015}. 
Environmental features~\cite{Bechinger2016}, such as geometric confinement~\cite{Wensnick2008,vedula2012emerging,wioland2013confinement,Wioland16,Duclos18,Kempf19,Jain20,Norton18,arora2024,canavello2024}, anisotropy of the embedding fluid~\cite{doi:10.1126/science.aah6936,Song2021}, structural details of the underlying substrate~\cite{He15,Gloag17,Nam2016,Lee2018,wen2023collective}, and obstacles~\cite{Kaiser2012},
further affect collective motion by reorganizing active flows or generating new dynamical states. In circular confinement, for example, many active systems spontaneously develop circulating flows that do not arise in unbounded environments~\cite{wen2023collective}.

How confinement influences the conversion of particle-scale propulsion dynamics into collective motion and boundary response remains unresolved.
This has led to increasing interest in active matter confined within flexible or deformable boundaries such as lipid vesicles and liquid droplets~\cite{tjhung12,giomi14,paoluzzi2016,chen2017,gao17,wang19,li2019,vutukuri2020,quillen20,peterson2021,ruske21,li2022,nagard22,kokot2022,lee2023,tiribocchi2023a,tiribocchi2023b,tiribocchi2023c,carlsson24}.  
These systems provide settings in which activity of the SPPs can reorganize or deform a soft enclosing structure, with examples ranging from synthetic platforms, such as active colloids or granular particles encapsulated in soft elastic rings, to biological membranes and active droplets. 
{\cbb In particular, the recent studies of active droplets and deformable active-fluid systems demonstrated that active stresses and confinement can generate spontaneous droplet motion, shape transitions, and migration through constrictions~\cite{tiribocchi2023a,tiribocchi2023b,tiribocchi2023c}. These studies highlight the important role of boundary mechanics and active-stress organization in regulating the dynamics of confined active systems.}
The interplay between active stresses and deformable boundaries is also central to many biological and biomimetic systems.
Many essential biological processes, such as mitosis, endocytosis, apoptosis, morphogenesis, and cell migration, stem from the dynamic responsiveness of the plasma membrane to active intracellular components~\cite{ananthakrishnan07,doherty09,needleman17}. 
The way in which active stresses couple to deformable boundaries is thus important to both biological physics and the design of synthetic active materials.

Studies examining SPPs under flexible confinement have focused largely on active Brownian particles, which propel with a nearly constant speed along an orientation set by their rotational diffusion~\cite{vutukuri2020,li2019,peng2022,li2022,chen2017,paoluzzi2016,wang19}. 
A common finding across these studies is that active Brownian particles tend to accumulate near the enclosure boundary and exert stresses that generate pronounced shape deformations. Although such models capture boundary accumulation and pressure generation, they do not incorporate alignment interactions, shape anisotropy, or 
environment--sensitive reorientation mechanisms. 
As a result, they cannot describe collective states in which particles interact, align, or adapt their propulsion in response to local cues, features that strongly influence collective phase behavior in many biological and engineered active systems.

Comparatively fewer investigations have addressed SPPs that interact with one another, align their direction of motion with neighboring particles, and are confined within flexible boundaries~\cite{quillen20,gao17,peterson2021}. Quillen {\it et al.}~\cite{quillen20} examined deterministic point--like SPPs confined by a 
two--dimensional semi--flexible ring and observed that low boundary bending rigidity permits particle--driven shape changes that produce sprocket--like geometries. 
Peterson {\it et al.}~\cite{peterson2021} studied semi--flexible polymer chains in three-dimensional enclosures, where each chain segment is propelled along its local tangent. They identified distinct spatiotemporal phases controlled by chain length and volume fraction. In particular, long chains accumulate near the boundary and form quasi-two-dimensional vortical flows, whereas shorter chains self--organize into multiple nematic caps whose number increases with chain length~\cite{peterson2021}.
Field-theoretical descriptions of active nematics in passive droplets have reported defect--driven flows and boundary deformations~\cite{tjhung12,giomi14,gao17,ruske21}.
While these studies reveal diverse collective states under flexible confinement, they leave open how microscopic propulsion rules influence collective dynamics and boundary response.

A biologically relevant class of SPPs consists of swimmers that reverse their direction of motion without tumbling.
These include the parasitic protozoan {\em Trypanosoma brucei}, which causes African trypanosomiasis and can reverse the chirality of its flagellar beat to move backward through densely obstructed vascular environments~\cite{heddergott2012trypanosome}; the polarly flagellated marine bacterium {\em Vibrio alginolyticus}, which employs a distinct reversal mechanism in aqueous media~\cite{xie2011bacterial}, in contrast to the 
run--and--tumble behavior characteristic of {\em Escherichia coli}~\cite{berg72}; and the myxobacterium {\em Myxococcus xanthus}, which can reverse its gliding direction to facilitate the outward expansion of cells at the edges of their swarms~\cite{wu2009periodic}. Such reversals introduce an additional intrinsic timescale that enables rapid reorganization of collective motion without loss of orientational order.
Although several computational studies have examined reversal--capable SPPs~\cite{Strefler2008,Grossmann2016-1,Grossmann2016-2,Olsen2021,wen2023collective},
their collective dynamics under soft confinement and the resulting coupling to boundary mechanics remain largely unexplored.

In this work, we investigate how flexible boundaries {reshape} the collective dynamics of reversal--capable SPPs using a simple coarse--grained simulation approach~\cite{Wen22}. The system consists of densely packed, environment--sensitive SPPs confined within deformable two--dimensional enclosures whose elastic moduli can be independently tuned. Each SPP is modeled as an elongated, disjoint ring polymer with two polar sites and is propelled by a motility force acting along its nematic polarity. 
The direction of this force is determined by the SPP’s recent average velocity computed over a tunable time window (memory time), giving rise to persistent motion and  
non--tumbling reversals.
This formulation couples particle--level reversibility, alignment interactions, and boundary mechanics, while maintaining independent control over activity and enclosure elasticity.
By varying the motility strength
and the duration of the velocity memory, we examine how boundary flexibility and SPP activity shape the collective dynamics and the resulting boundary morphology.
{\cbb Our model is intentionally minimal: it does not include explicit hydrodynamic interactions, incompressibility constraints, adhesion to the boundary, or strict area conservation of the enclosure. As a result, the deformable boundary should not be interpreted as a quantitative model of a lipid vesicle or active fluid droplet. Instead, it provides a generic framework for isolating how boundary deformability and particle-level memory effects reorganize collective active motion. While these additional physical ingredients may quantitatively affect stress propagation, shape deformations, and propulsion dynamics, we expect the qualitative interplay between active stresses, collective organization, and boundary mechanics identified here to persist qualitatively.}

We find that reversal--capable SPPs confined within deformable enclosures display four distinct dynamical regimes as activity increases: disordered motion, reversal-rich vortical flow, persistent unidirectional vortical flow, and a high-activity ballistic state in which global polar alignment drives sustained translation of the entire enclosure. 
Boundary flexibility plays an important role in organizing these states: it substantially broadens the activity range over which vortical and polar motions coexist, strongly enhances vorticity-reversal rates, and shifts the onset of persistent unidirectional vortical flow to higher motility strengths.
Comparisons with matched-area rigid enclosures confirm that these effects originate from boundary flexibility rather than from the increase in enclosure area induced by SPP stresses.

\section{Model and Method} \label{MM}

We consider a system of ${\cal P}$ disjoint SPPs that move in two-dimensional space. Each SPP is modeled as a semi--flexible ring polymer composed of $N$ beads, in an implicit good solvent~\cite{zhu21}. We recently employed this model to explore the collective behavior of environment--sensitive SPPs on uniform and circularly patterned substrates~\cite{Wen22,wen2023collective}. In this model, an SPP is elongated and experiences a self--propulsion force acting along its long axis. The SPPs are confined by a flexible boundary, modeled as a large passive ring polymer.  
The total potential energy of the system consists of three main components: the net potential energy of the SPPs, ${\cal U}_{S}$; 
the potential energy of the flexible confining boundary, ${\cal U}_{C}$; and the interaction potential energy between the SPPs and the boundary, ${\cal U}_{SC}$.
The SPPs' potential energy is given by 
\begin{eqnarray}\label{eq:net_potential}
{\cal U}_{S}\!&=& \! \sum_{l=1}^{\cal P}\!\left[\!\sum_{i=1}^N{\cal U}_{S,bond}\!\left(r^{(l)}_{i,i+1}\right)
\!\!+\!\! \sum_{i=1}^N{\cal U}_{S,bend}\!\left(\!{ \bf r}^{(l)}_{i-1}\!,\!{\bf r}^{(l)}_{i}\!,\!{\bf r}^{(l)}_{i+1}\right)\!\!\right]\nonumber \\
 &+&\! \sum_{l,k}\sum_{i, j}{\cal U}_{S,rep}\!\left(|{\bf r}^{(l)}_{i}-{\bf r}^{(k)}_{j} |\right)\!\!+\!\! 
 \sum_{l=1}^{\cal P}{\cal U}_{S,area}\!\left(\left\{\rr^{(l)}_{i}\right\}\!\right)\!,
\end{eqnarray}
with $r^{(l)}_{i,j}=|{\bf r}^{(l)}_i-{\bf r}^{(l)}_j|$, where $\rr^{(l)}_i$ denotes the position of bead $i$ belonging to SPP $l$, satisfying the boundary conditions ${\bf r}^{(l)}_{N+1}={\bf r}^{(l)}_1$ and ${\bf r}^{(l)}_{0}={\bf r}^{(l)}_N$.
In Eq.~(\ref{eq:net_potential}),
${\cal U}_{bond}$ represents a harmonic potential that maintains the connectivity between consecutive beads within an SPP and reads
\BE \label{eq:bonding}
{\cal U}_{S,bond}(r)=\frac{k_{S}}{2}(r-l_S)^2,
\EE
where $k_{S}$ is the spring constant of the rings and $l_S$ denotes the preferred bond length.
${\cal U}_{S,bend}$, in Eq.~(\ref{eq:net_potential}), 
is a three--body potential that enforces the bending stiffness of the SPPs, with a preferred bond angle of $\theta_0=\pi$.
It is given by 
\BE \label{eq:bending}
{\cal U}_{S,bend}\left({\bf}r^{(l)}_{i-1},{\bf}r^{(l)}_{i},{\bf}r^{(l)}_{i+1}\right)=\kappa_S(1+\cos\theta),
\EE
where $\cos\theta={\bf r}_{i-1,i}^{(l)}{\bf \cdot} {\bf r}^{(l)}_{i+1,i}/{r}_{i-1,i}^{(l)}{r}^{(l)}_{i+1,i}$ and $\kappa_S$ denotes the bending stiffness of the SPP rings. 

To introduce geometric anisotropy into the SPPs, monomer triplets centered at $i=1$ and $i=N/2+1$ are assigned a preferred bond angle $\theta_p<\pi$. 
At these polar sites, the following three--body bending potential is applied
\BE \label{eq:bending2}
{\cal U}_{S,bend}\left({\bf}r^{(l)}_{i-1},{\bf}r^{(l)}_{i},{\bf}r^{(l)}_{i+1}\right)=
\frac{\kappa'_S}{2}(\cos\theta-\cos\theta_p)^2.
\EE
This potential is softer near the preferred angle than the standard bending interaction defined in Eq.~(\ref{eq:bending}). 
We found that choosing $\kappa'_S \approx 10\kappa_S$ provides an equivalent level of stiffness at the poles as elsewhere along the ring.

A good solvent condition for the ring polymers is enforced by introducing a repulsive interaction between any two non--consecutive monomers, defined by the potential
\begin{eqnarray} \label{eq:SPP-repulsive}
{\cal U}_{S,rep}(r)=
\left\{\begin{array}{cc}
\frac{\zeta_{SS}}{2}\left(1-\frac{r}{r_c}\right)^2 &{ \mathrm {if\ } r\leq r_c,} \\
{0} &{\mathrm{if}}\  \  r>r_c, \end{array}\right. 
\end{eqnarray}
where $\zeta_{SS}$ and $r_c$ denote the strength and range of the repulsive interaction, respectively. For sufficiently large $\zeta_{SS}$, this interaction, together with an initial configuration in which the rings are non--overlapping, maintains the disjoint nature of the SPPs throughout the simulation.

Each SPP is also subject to an area constraint, enforced 
through the potential energy
\BE \label{eq:area_constraint}
{\cal U}_{S,area}\left(\{\rr^{(l)}_{i}\}\right)=\frac{\chi_S}{2}\left(A_S-A\left(\{\rr^{(l)}_{i}\}\right)\right)^2,
\EE
where $\chi_S$ is the area--stretch modulus, $A_S$ is the preferred enclosed area of the SPP, and $A$ is the actual enclosed area, computed using the shoelace formula:
\BE\label{eq:shoelace}
A\left(\{\rr^{(l)}_{i}\}\right)=\frac{1}{2}\biggl{|}\sum_{i=1}^N\left(x^{(l)}_iy^{(l)}_{i+1}-x^{(l)}_{i+1}y^{(l)}_{i}\right)\biggl{|}.
\EE

The SPPs are confined within a boundary that is also modeled as a semi--flexible ring polymer consisting of $M$ beads. The potential energy of this boundary is given by
\BE\label{eq:enclosure_energy}
{\cal U}_C=\sum_{i=1}^M\left[{\cal U}_{C,bond}\left(R_{i,i+1}\right)+{\cal U}_{C,bend}\left({ \bf R}_{i-1},{\bf R}_{i},{\bf R}_{i+1}\right)\right],
\EE
where ${\bf R}_i$ denotes the position of bead $i$ belonging to the confining boundary and $R_{i,j}=|{\bf R}_i-{\bf R}_j|$. The bond potential ${\cal U}_{C,bond}$ follows Eq.~(\ref{eq:bonding}) with spring constant $k_{C}$ and preferred bond length $l_C$. 
Similarly, the bending potential ${\cal U}_{C,bend}$ follows Eq.~(\ref{eq:bending}) with bending stiffness $\kappa_C$ and preferred bond angle $\pi$. Confinement of the SPPs is maintained by a repulsive interaction ${\cal U}_{SC}$, which has the same functional form as Eq.~(\ref{eq:SPP-repulsive}), with interaction strength $\zeta_{SC}$ and range $r_c$.

In this model, each SPP $l$ is propelled by a motility force of magnitude $F_D$, which acts either parallel or antiparallel to its polarity.  This force is defined as~\cite{kabla12}
\begin{equation} \label{eq:motility_force}
{\bf f}_l(t) = F_D\,\,  g\left(\bar{\bf v}_l(t),{\bf\hat n}_l(t)\right){\bf\hat n}_l(t),
\end{equation}
where the unit vector 
\BE\label{eq:unit_vector}
{\bf\hat n}_l(t)=\left[{\bf r}^{(l)}_{N/2+1}(t)-{\bf r}^{(l)}_{1}(t)\right]/\left|{\bf r}^{(l)}_{N/2+1}(t)-{\bf r}^{(l)}_{1}(t)\right|
\EE
represents the polarity of SPP $l$. The function $g({\bf A},{\bf B})$, which takes the value $+1$ if ${\bf A}\cdot{\bf B} >0$ or $-1$ if ${\bf A}\cdot{\bf B}<0$,  determines the orientation of the motility force based on its recent direction of motion, quantified by its average velocity $\bar{\bf v}_l(t)$ over the time interval $\left[t-\tau_m,t\right]$, i.e., 
\BE\label{eq:historical_velocity}
\bar{\bf v}_l(t)= \frac{1}{\tau_{m}}\int_{t-\tau_{m}}^t{\bf v}_l(t')dt'. 
\EE
This formulation implies that, although the SPPs are nematic, they are not intrinsically polar and can reverse their direction without a tumbling event.  
Henceforth, $\tau_m$ will be referred to as the persistence time.
{\cbb For low values of the driving force, thermal fluctuations dominate the dynamics, leading to a high reversal rate of isolated SPPs in the dilute regime. As $F_D$ increases, the reversal rate decreases rapidly and, for the parameters used in the present study with $\tau_m=1\tau$, reversals become essentially absent for $F_D\gtrsim 8 \varepsilon/r_c$, as shown in Fig.~S1 of the ESI. Representative trajectories of isolated SPPs in the dilute regime at low and high values of $F_D$ are shown in Fig.~S2 of the ESI.}

The monomers of the SPPs move according to a molecular dynamics scheme with a Langevin thermostat,
\begin{eqnarray} 
	\dot{\bf r}^{(l)}_{i}(t) &=& {\bf v}^{(l)}_{i}(t),\label{eq:velocity}\\
	m_S\dot{\bf v}^{(l)}_{i}(t)& =& -\nabla^{(l)}_{i}{\cal U}+\frac{1}{N}{\bf f}_l(t)-\gamma\,{ \bf v}^{(l)}_{i}(t)\nonumber \\
    &+&\sqrt{2\gamma k_BT} \, {\boldsymbol\xi}^{(l)}_i(t), \label{eq:acceleration}
\end{eqnarray}
where ${\cal U}={\cal U}_S+{\cal U}_C+{\cal U}_{SC}$ is the net potential energy of the system,
$m_S$ is the mass of a single bead, $\gamma$ is the friction coefficient, $k_B$ is Boltzmann's constant, and $T$ is temperature.  
The operator $\nabla^{(l)}_i=\left(\partial/\partial x_i^{(l)},\partial/\partial y_i^{(l)}\right)$.  
${\boldsymbol\xi}_i^{(l)}$ denotes 
a random vector drawn from a uniform distribution~\cite{greiner88,duenweg91} whose moments satisfy
\begin{eqnarray}
\langle{\boldsymbol\xi}^{(l)}_{i}(t)\rangle&=&0\nonumber\\
\langle{\xi}^{(k)}_{i,{\mu}}\left(t\right){\xi}^{(l)}_{j,{\nu}}\left(t'\right)\rangle & = &\delta_{kl}\delta_{\mu\nu} \delta_{ij}\delta\left(t-t'\right),
\end{eqnarray}
where $\mu$ and $\nu$ correspond to the Cartesian coordinates $x$ or $y$. Numerically, ${\xi}^{(l)}_{i,\mu}\in \left[-\sqrt{3/\Delta t},\sqrt{3/\Delta t}\,\right]$, 
where $\Delta t$ is the integration time step.

Since the enclosing boundary is passive, its beads obey the following equations of motion
\begin{eqnarray} 
	\dot{\bf R}_{i}(t) &=&  {\bf V}_{i}(t),\label{eq:velocity2}\\
	 m_C\dot{\bf V}_{i}(t)& =& -\nabla_{i}{\cal U}-\Gamma\,{ \bf V}_{i}(t)+
    \sqrt{2\Gamma k_BT}\, \boldsymbol{\Xi}_i(t)
    , \label{eq:acceleration2}
\end{eqnarray}
where $m_C$ is the mass of a single bead of the confining boundary, $\Gamma$ is its friction coefficient, and $\boldsymbol{\Xi}_i$ is a random vector that obeys the same conditions as those of the SPP beads.

In our model, each SPP is composed of $N=40$ beads, and the confining boundary is composed of $M=942$ beads. The values of the parameters that are kept fixed are as follows
\begin{eqnarray}\label{eq:parameters}
 k_S&=&100\,\varepsilon/r_c^2,\, \kappa_S=100\,\varepsilon,\,\kappa'_S=1000\,\varepsilon,\, \theta_p=120^{\circ}, \nonumber\\
 \zeta_{SS}&=&50\,\varepsilon,\, \zeta_{SC}=100\,\varepsilon, \, \chi_S=1\,\varepsilon/r_c^2,\, A_S=100\,r_c^2,\nonumber\\
 l_S&=&r_c, \,  l_C=0.5\,r_c, \,  \kappa_C=100\,\varepsilon, \, k_BT=\varepsilon, \nonumber\\
 \gamma&=&1.0\,m_S/\tau, 
 \, \Gamma=1.0\,m_C/\tau,\, {\rm and }\,\, m_C=m_S,
\end{eqnarray}
where the timescale $\tau=r_c\sqrt{m_S/\varepsilon}$.
The effects of the elasticity of the confining boundary are examined by varying the bonding coefficient $k_C$ from $200\,\varepsilon/r_c^2$ to $4000\,\varepsilon/r_c^2$.
In most simulations, however, $k_C=2000\,\varepsilon/r_c^2$.
The magnitude of the motility force $F_D$ is varied between 0 and $64\,\varepsilon/r_c$, corresponding to a P{\'eclet} number, ${\rm Pe}=LF_D/4k_BT$ ranging between 0 and about 225, where $L\approx 14\, r_c$ is the length of an SPP.
Hereafter, times and lengths are expressed in units of $\tau$ and $r_c$, respectively. 
Equations~(\ref{eq:velocity}-\ref{eq:acceleration2}) are integrated numerically using the velocity--Verlet method~\cite{swope82} with an integration time step $\Delta t=0.01\,\tau$. Most simulations are run over $5\times 10^7$ time steps. In what follows we fix $\tau_m=1.0\,\tau $ unless otherwise mentioned.

{\cbb A schematic of a single SPP and its propulsion mechanism is shown in Fig.~S2(A) of the ESI.
To characterize the motility of an isolated SPP, we performed simulations in the dilute
regime in the absence of confinement. Representative trajectories of a single SPP at
different values of $F_D$ are shown in Fig.~S2(B). At low $F_D$, the SPP undergoes
diffusive-like motion with frequent reversals, resulting in compact, localized trajectories.
As $F_D$ increases, reversals are progressively suppressed and the SPP explores
increasingly large regions of space, with the trajectory becoming more persistent and
extended. The mean-square displacement $\Delta R^2_{SPP}(t)$ of the SPP center of mass,
shown in Fig.~S2(C), exhibits a crossover from ballistic behavior at short times
($\Delta R^2_{SPP} \sim t^2$) to diffusive behavior at long times ($\Delta R^2_{SPP} \sim t$),
consistent with the behavior expected for a self-propelled particle with finite persistence.
The crossover timescale increases markedly with $F_D$, reflecting the suppression of
reversals at higher driving forces and the consequent enhancement of motional persistence.
At $F_D = 0$, the short-time ballistic regime arises from the inertial dynamics of the
Langevin thermostat rather than from activity.}

We quantify the collective dynamics of the SPPs using two order parameters: the vortical and Vicsek order parameters. 
The vortical order parameter, which quantifies the degree of coherent rotational motion, is defined by
\begin{equation}
S_{V}(t)=\biggl{|}\frac{1}{\cal P}\sum_{l=1}^{\cal P}{\sigma_{l}(t)}\biggl{|}, 
\label{eq:vortical-order}
\end{equation}
where $\sigma_{l} (t)= +1$ or $-1$ when the tangential velocity of SPP $l$ is directed clockwise or counterclockwise, respectively. The tangential velocity is computed with respect to the center of mass of the confining boundary. The Vicsek order parameter, which measures the global alignment of  SPP velocities, is defined by
\begin{equation}
S_{N}(t)=\biggl{|}\frac{1}{\cal P}\sum_{l=1}^{\cal P}{e^{i\alpha_{l}(t)}}\biggl{|},
\label{eq:vicsek-order}
\end{equation}
where $\alpha_{l}(t)$ is the instantaneous angle between the velocity of SPP $l$, $\textbf{v}_l(t)=(1/N)\sum_{i=1}^N{\bf v}_i^{(l)}(t)$, and a fixed reference axis (e.g., the $x$--axis). 

\begin{figure}[t]
  \begin{center}
	\includegraphics[scale=0.539]{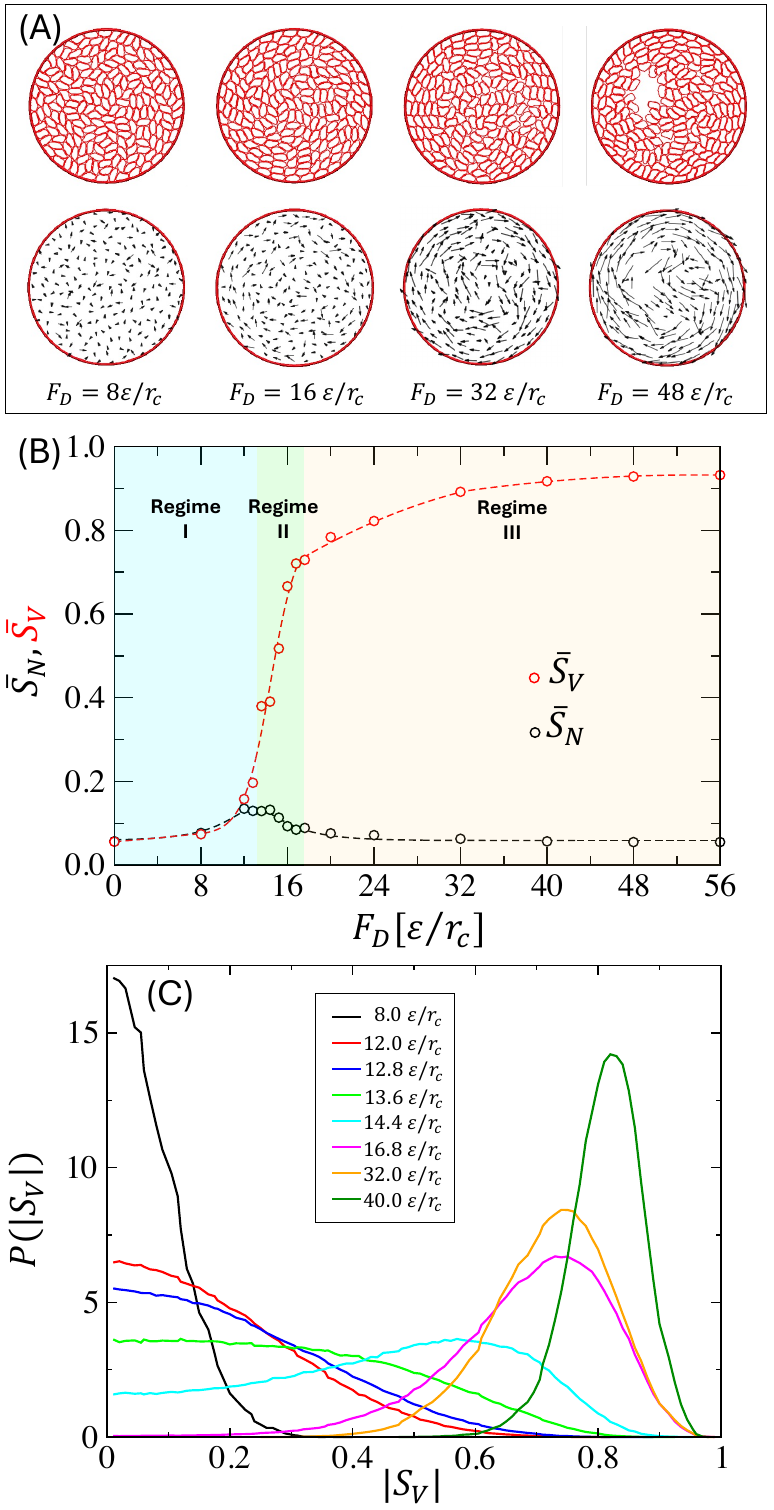}
   \end{center}
\caption{ 
(A) Configuration snapshots (top row) and corresponding velocity fields (bottom row) of SPPs confined within a rigid enclosure of radius $a=90\,r_c$ at different motility forces $F_D$. (B) Vicsek order parameter $\bar S_N$ and vortical order parameter $\bar S_V$ as functions of $F_D$. The three dynamical regimes (I--III) are described in the main text.
(C) Normalized distribution of the absolute value of the instantaneous vortical order parameter, $|S_V|$, for different values of $F_D$.}
\label{fig:snapshots_order_param_rigid}
\end{figure}

\section{Results}

To establish a baseline for comparison with flexible confinement, we first simulated reversal--capable SPPs within a rigid circular boundary. Confinement was enforced through the repulsive interaction described by Eq.~(\ref{eq:SPP-repulsive}) with the interaction strength set to $\zeta_{SC}$ as specified in Eq.~(\ref{eq:parameters}). The beads of the confining boundary were held fixed.  The motility force $F_D$ was systematically varied while maintaining a packing fraction $\phi = {\cal P}A_S/\pi a^2 = 0.786$, where $a$ is the radius of the rigid enclosure.

The time--averaged vortical order parameter, ${\bar S}_V$, computed from Eq.~(\ref{eq:vortical-order}) in steady state and shown in Fig.~\ref{fig:snapshots_order_param_rigid}(B), exhibits a sharp increase at $F_D\gtrsim 13\,\varepsilon/r_c$. This rise marks the onset of coherent vortical motion of the SPPs under rigid confinement and is consistent with previous findings~\cite{wen2023collective,wioland2013confinement,zuiden2016}. Movies 1 and 2 in the ESI illustrate representative dynamics at $F_D=8\,\varepsilon/r_c$ and $16\,\varepsilon/r_c$, respectively. 

\begin{figure}[t!]
  \begin{center}
	\includegraphics[scale=0.55]{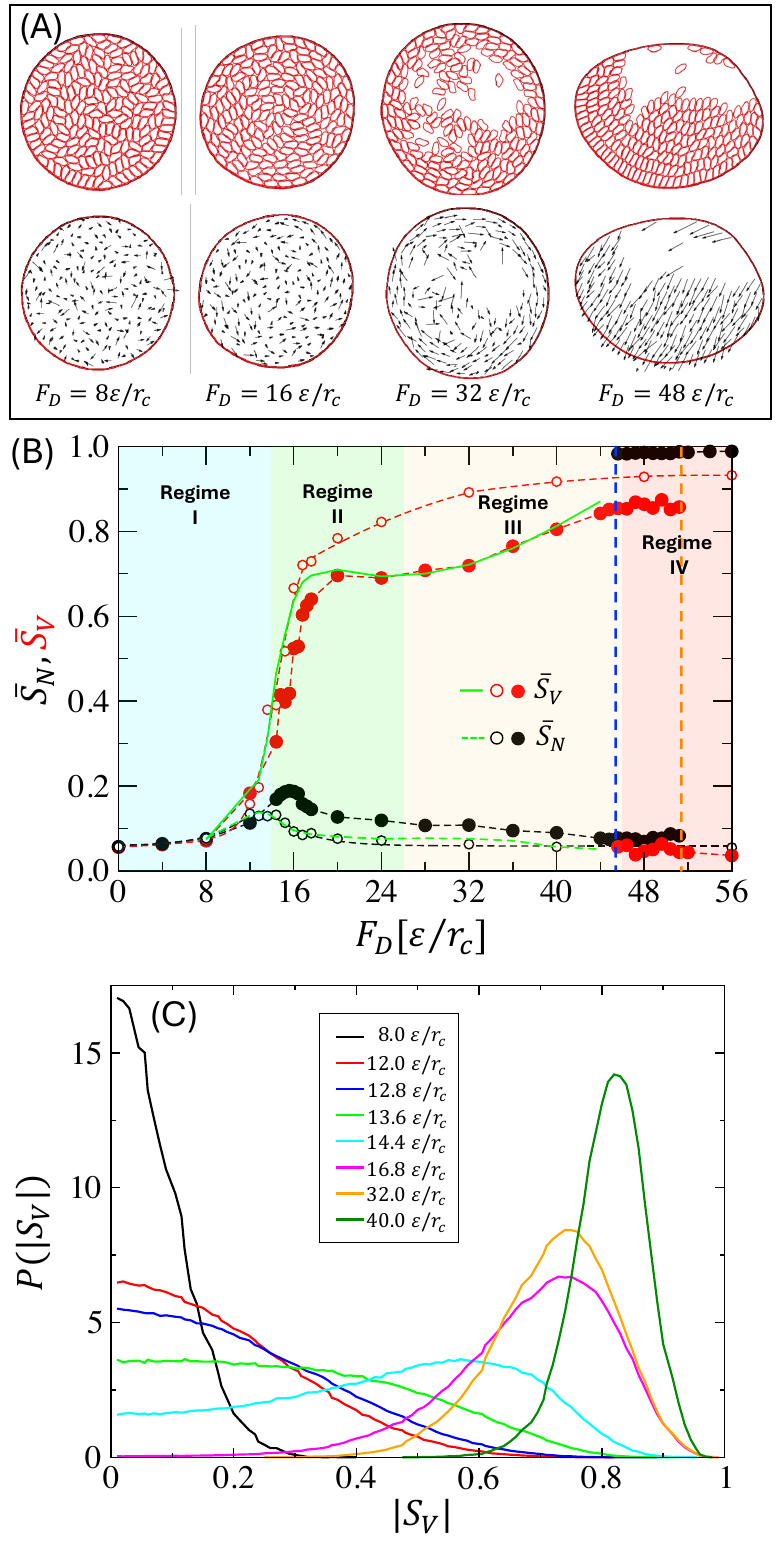}
   \end{center}
\caption{(A) Configuration snapshots (top row) and corresponding velocity fields (bottom row) in the case of a flexible enclosure with $k_C = 2000\,\varepsilon/r_c^2$ shown at different motility forces.
(B) Vicsek (${\bar S}_N$) and vortical (${\bar S}_V$) order parameters vs. $F_D$. Solid and open circles correspond to the case of flexible and rigid boundaries (with $a = 90\,r_c$), respectively, as also shown in Fig.~\ref{fig:snapshots_order_param_rigid}. Solid and dashed green lines denote ${\bar S}_V$ and ${\bar S}_N$, respectively, for the rigid--boundary case with the enclosure area fixed to the average area of the flexible enclosure at the same $F_D$. The four dynamical regimes, discussed in the text, are indicated by shaded regions. The vertical dashed blue line marks the extent of bistability interval of the ballistic regime (Regime IV) within the vortical regime (Regime III), while the vertical dashed orange line marks the extent of bistability of the vortical regime within the ballistic regime.
(C) Normalized distribution of the absolute value of the instantaneous vortical order parameter, $|S_V|$, for different values of $F_D$.}
\label{fig:snapshots_order_param_flexible}
\end{figure}

The time--averaged Vicsek order
parameter ${\bar S}_N$ remains low across all values of $F_D$, consistent with the high packing fraction and rigid boundary, which restrict large--scale translational motion of the SPPs.
A modest increase in ${\bar S}_N$ is nevertheless observed over the same range of $F_D$ in which ${\bar S}_V$ rises sharply.
This enhancement arises from the intrinsic flexibility of the SPPs, which introduces local compressibility and permits small localised rearrangements, enabling weak nematic ordering and collective polar motion of the SPPs under rigid confinement.

The emergence of vortical motion at $F_D\approx 13\,\varepsilon/r_c$ is evident from the peak of the distribution of the instantaneous vortical order parameter, $P(|S_V|)$, shifting from zero to a finite value, as shown in Fig.~\ref{fig:snapshots_order_param_rigid}(C). In this regime, the direction of vortical motion switches stochastically, 
with a reversal rate that decreases with increasing $F_D$, as will be discussed below.  
For $F_D\gtrsim 17\,\varepsilon/r_c$, 
the system enters a state of persistent unidirectional vortical motion: the distribution $P(|S_V|)$ becomes sharply peaked away from zero, while the probability of small $|S_V|$ values vanishes, indicating suppression of reversal events.
Accordingly, the collective behavior of SPPs under rigid confinement exhibits three distinct dynamical regimes: Regime I (low $F_D$) characterized by noncoherent motion of the SPPs; Regime II (intermediate $F_D$) characterized by collective vortical motion with intermittent reversals and weak polar motion; and Regime III (high $F_D$) characterized by persistent unidirectional vortical motion with negligible
polar motion.

To examine how boundary flexibility affects the collective dynamics, we next performed simulations in which the SPPs were confined within a flexible enclosure prepared as described in Sec.~\ref{MM}, with elastic parameters corresponding to a stretch modulus of $k_C = 2000\,\varepsilon/r_c^2$ and a bending modulus of $\kappa_C = 100\,\varepsilon$.
The resulting time--averaged vortical and Vicsek order parameters, together with representative snapshots, are shown in Fig.~\ref{fig:snapshots_order_param_flexible} (red and black solid circles, respectively).
For comparison, Fig.~\ref{fig:snapshots_order_param_flexible}(B) also includes the vortical and Vicsek order parameters for the case of rigid confinement with $a = 90\, r_c$, already presented in Fig.~\ref{fig:snapshots_order_param_rigid}(B) (red and black open circles, respectively).

Figure~\ref{fig:snapshots_order_param_flexible}(B) shows that 
boundary deformability significantly modifies the collective dynamics of the SPPs relative to rigid confinement.
Most notably, flexible confinement gives rise to a fourth high--activity regime (Regime IV), in which 
global polar alignment of the SPPs generates sustained ballistic motion of the entire enclosure,
as illustrated in Movie 3 of the ESI.
Boundary deformability also substantially broadens Regime II, characterized by 
reversal--rich vortical flow, expanding its range to $14\, \varepsilon/r_{\rm c}\lesssim F_{\rm D}\lesssim 26\,\varepsilon/r_{\rm c}$ from only $13\, \varepsilon/r_{\rm c}\lesssim F_{\rm D}\lesssim 17\, \varepsilon/r_{\rm c}$ under rigid confinement. 
In addition, boundary deformability 
leads to a pronounced enhancement of the vorticity--reversal rate within Regime II, as analyzed below. 

The motile SPPs exert outward pressure on the confining boundary, leading to an increase in the enclosure area with increasing $F_D$, as shown in Fig.~{\cbb S3} of the ESI. Notably, the area increase is gradual across Regimes I--III and does not exhibit sharp changes at the crossovers between these regimes. This observation raises the question of whether the qualitative differences observed between rigid and flexible confinement arise primarily from reduced effective packing fraction due to vesicle expansion, or instead from enclosure flexibility itself, which we address below.

To distinguish the effect of reduced packing from that of boundary flexibility,
we performed additional simulations with rigid circular confinement whose area was matched to the average area of the flexible enclosure at the same motility force. The resulting time--averaged vortical and Vicsek order parameters are shown in Fig.~\ref{fig:snapshots_order_param_flexible}(B) as solid and dashed green lines, respectively. 
The two order parameters of the matched--area system closely follow those of the 
fixed--radius case ($a = 90\,r_c$) up to $F_D \approx 16\,\varepsilon/r_c$. This demonstrates that the shift in the onset of Regime II originates from boundary flexibility rather than from the increase in enclosure area.
{\cbb Figures~\ref{fig:snapshots_order_param_rigid}(A) and~\ref{fig:snapshots_order_param_flexible}(A) show that at low activity (Regime I and the early part of Regime II), the SPPs remain nearly confluent, whereas at larger $F_D$ the increase in enclosure area leads to significant void formation and non-confluent configurations. As such, the collective behaviors discussed in the later regimes are not associated with confluence effects.}

\begin{figure}[t]
  \begin{center}
	\includegraphics[scale=0.67]{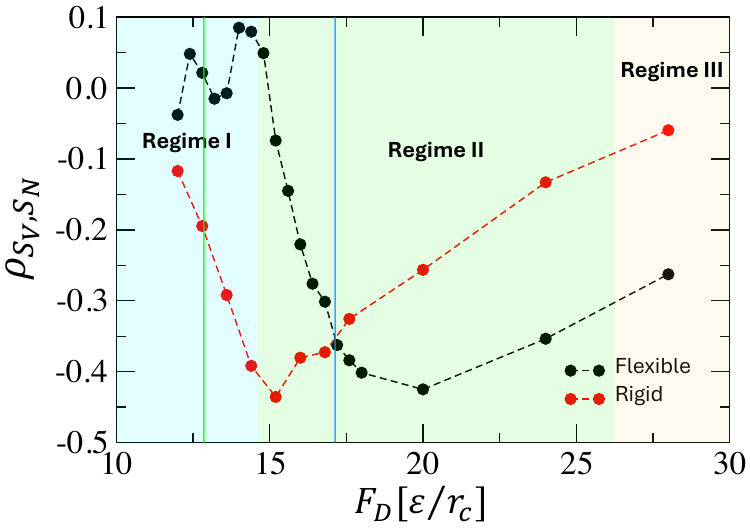}
   \end{center}
\caption{Pearson correlation $\rho_{S_V,S_N}$ of the vortical and Vicsek order parameters vs. motility force for the case of flexible enclosure with $k_C=2000\,\varepsilon/r_c^2$ (black points), and for rigid enclosure (red points). The shaded regions correspond to Regimes I--III in the flexible--confinement case. The vertical green (blue) line separates Regime I (II) from Regime II (III) in the rigid--confinement case. }
\label{fig:pearson_order_parameters}
\end{figure}

To further investigate the collective behavior of the SPPs in Regime II, we computed the
Pearson correlation coefficient $\rho_{S_V,S_N}$ between the vortical and Vicsek order parameters for both rigid and flexible confinement, defined as 
\BE \label{eq:pearson}
\rho_{S_V,S_N}=\frac{\langle \delta S_V(t)\delta S_N(t)\rangle_t}{\sqrt{\langle\delta S_V^2(t)\rangle_t\langle\delta S_N^2(t)}\rangle_t},
\EE
where $\delta S_V(t)=S_V(t)-{\bar S}_V$ and $\delta S_N(t)=S_N(t)-{\bar S}_N$. Figure~\ref{fig:pearson_order_parameters} shows that $\rho_{S_V,S_N}$ is weak in Regime I, but becomes strongly negative in Regime II, reaching its lowest values in the range of $F_D$ where ${\bar S}_N$ is relatively high and ${\bar S}_V$ rises sharply. 
Because $\rho_{S_V,S_N}$ is computed from instantaneous fluctuations, 
these strongly negative values indicate a clear dynamical anticorrelation between vortical and polar collective motion: episodes of enhanced polar motion occur preferentially when vortical motion is suppressed, and vice versa.

{\cbb
We now turn to Regime IV, characterized by ballistic collective motion of the SPPs that emerges at sufficiently high motility forces, appearing at $F_D\approx 46\,\varepsilon/r_c$ for $k_C=2000\,\varepsilon/r_c^2$ and $\kappa_C=100\,\varepsilon$. This regime is marked by global alignment of both the polarities and velocities of the SPPs, together with persistent ballistic motion of the enclosure. As shown in Fig.~\ref{fig:snapshots_order_param_flexible}(A) and Movie 3 in the ESI, the SPPs accumulate at the leading edge of the enclosure, while the boundary elongates perpendicular to the average velocity of the SPPs.

The transition from Regime III to Regime IV is discontinuous in character, with a 
bistability region in which the system can settle into either a long-lived vortical 
or ballistic state depending on initial conditions. To map this region, we performed 
10 independent simulations at each of 11 values of $F_D$ ranging from 
$44.8\,\varepsilon/r_c$ to $56\,\varepsilon/r_c$, spanning from well within the 
vortical regime to well within the ballistic regime. At $F_D = 44.8\,\varepsilon/r_c$ 
all 10 simulations reached the vortical state, while for $F_D \geq 52\,\varepsilon/r_c$ 
all reached the ballistic state. Within the interval 
$45.6\,\varepsilon/r_c \lesssim F_D \lesssim 51.2\,\varepsilon/r_c$ (between the 
vertical blue and orange dashed lines in 
Fig.~\ref{fig:snapshots_order_param_flexible}(B)), the probability of reaching the 
ballistic state increased monotonically with $F_D$ (see Table~S1 in the ESI). In 
some cases the system initially displayed vortical motion before spontaneously 
transitioning to the ballistic state, while the reverse transition was never observed. 
When $F_D$ was decreased from the ballistic regime, the system returned to vortical 
motion only at substantially lower values of $F_D$, confirming hysteretic behavior 
characteristic of a discontinuous dynamical transition.

}

\section{Discussion}

From the results described above, we see that boundary flexibility has a profound impact on the collective motion of the environment--sensitive SPPs.  We will now analyze this in more detail. 

\begin{figure}[t]
  \begin{center}
	\includegraphics[scale=0.49]{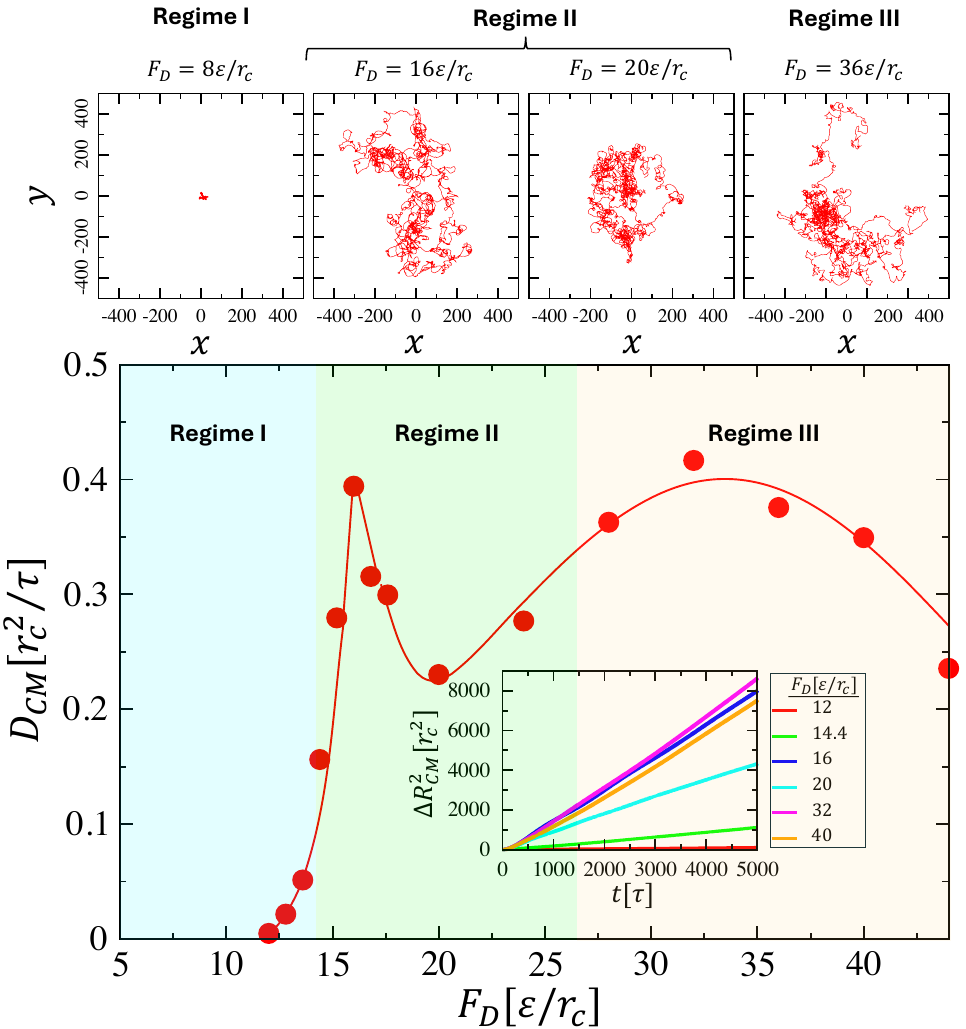}
   \end{center}
\caption{(Top) Trajectories of the enclosure's center of mass at different values of $F_D$. (Bottom) Diffusivity of the enclosure's center of mass vs. $F_D$. The red solid line is just a guide to the eye. Inset: Mean--square displacement of the enclosure's center of mass vs. time for different values of $F_D$.
Results correspond to a flexible enclosure with  $k_C=2000\,\varepsilon/r_c^2$.}
\label{fig:diffusivity}
\end{figure}

To explore the coupling between SPPs motion and boundary fluctuations, we first characterize the effect of SPP collective motion on the enclosure by analyzing the global mean--square displacement (MSD) of its center of mass, defined as $\langle \Delta R^2_{CM}(t)\rangle=\langle | {\bf R}_{CM}(t+t_0)-{\bf R}_{CM}(t_0)|^2\rangle_{t_0}$, shown in the inset of Fig.~\ref{fig:diffusivity} for different values of $F_D$. 
The long--time motion of the enclosure in Regimes I--III is diffusive, as evidenced by the linear growth of of its center--of--mass MSD (inset of Fig.~\ref{fig:diffusivity}). The corresponding diffusivity of the enclosure, defined as $D_{CM}=\langle \Delta R^2_{CM}(t)\rangle/4t$, shown as a function of $F_D$ in
Fig.~\ref{fig:diffusivity}, reveals 
distinct regimes of enclosure diffusion. The trajectories of the center of mass of the enclosure, in the {\cbb three} regimes, are shown in the top panel of  Fig.~\ref{fig:diffusivity}.

In Regime I, the enclosure exhibits very weak diffusion (Movie 4). Here, SPP motion is dominated by thermal fluctuations, resulting in largely uncorrelated forces that mostly cancel, thereby producing only a weak instantaneous net force on the enclosure.

In Regime II, the system does not exhibit a single mode of motion. Instead, transient episodes of coherent polar motion of the SPPs arise during transitions between clockwise and counterclockwise vortical states. This behavior is most pronounced near
$F_D\approx 16\,\varepsilon/r_c$ (Movie 5), where the Vicsek order parameter reaches its maximum.
During these short--lived events of coherent, albeit weak, polar motion, the SPPs generate a finite net force on the boundary that is larger than in Regime I, producing directed motion over finite time intervals. Frequent vorticity reversals repeatedly reorient this force, so that successive propulsion events are only weakly correlated in direction. As a result, the long--time motion remains diffusive, but with enhanced diffusivity due to the finite amplitude of these short--lived net forces.  
As $F_D$ is increased further, vorticity reversals are progressively suppressed, leading to a decrease in the enclosure  diffusivity. At still larger driving forces (Regime III), the diffusivity again increases and exhibits a broad maximum around $F_D\approx 32\,k_BT/r_c$. The origin of this peak is discussed later.

{\cbb 
To characterize the motion of individual SPPs within the confined ensemble, and to contrast it 
with the diffusive motion of the enclosure discussed above, we show in Fig.~S4 the trajectory 
and mean-square displacement of a single SPP. Panel~(A) shows representative trajectories at 
$F_D = 8$, $16$, and $32\,\varepsilon/r_c$, revealing the rotational character of the motion 
and the progressive localization of the SPP near the boundary with increasing $F_D$. 
Panel~(B) shows the corresponding mean-square displacement $\Delta R^2_\mathrm{SPP}(t)$. 
At short to intermediate times, individual SPP motion is ballistic 
($\Delta R^2_\mathrm{SPP} \sim t^2$), in sharp contrast to the diffusive motion of the 
enclosure center of mass. At late times, $\Delta R^2_\mathrm{SPP}(t)$ approaches but does 
not yet reach slope~1 within the simulation window. This is understood from the decomposition 
$\mathbf{R}_\mathrm{SPP}(t) = \mathbf{R}_{CM}(t) + \mathbf{r}_\mathrm{SPP}(t)$, 
which gives
\begin{equation}
\Delta R^2_\mathrm{SPP}(t) = 4D_{CM}t + C,
\end{equation}
where $C = 2\langle r^2_\mathrm{SPP}\rangle \sim a^2 \approx 8100\,r_c^2$ is the variance 
of the SPP position relative to the enclosure center of mass, with $a \approx 90\,r_c$ 
the enclosure radius and $D_{CM} \approx 0.3\,r_c^2/\tau$ the enclosure diffusivity (see Fig.~\ref{fig:diffusivity}). 
Since the crossover to slope~1 requires $t \gg a^2/4D_{CM} \approx 6750\,\tau$, and $C$ 
remains about $70\%$ of $4D_{CM}t$ at $t = 10^4\,\tau$, the long-time diffusion of the 
individual SPP is only beginning to become slaved to that of the enclosure within 
the accessible simulation time.
}

\begin{figure}[t]
  \begin{center}
	\includegraphics[scale=0.67]{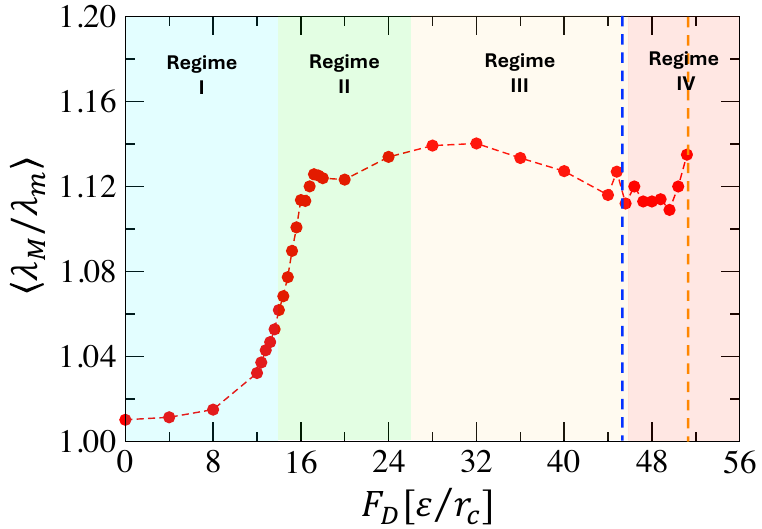}
   \end{center}
\caption{Shape anisotropy, defined as the ratio $\langle\lambda_M/\lambda_m\rangle$, from the gyration tensor in Eq.~(\ref{eq:gyration_tensor}), as a function of the driving force $F_D$ for a flexible enclosure with $k_C=2000\varepsilon/r_c^2$. Data shown in Regime IV are extracted from runs in which SPP motion is vortical. 
In the runs where the SPPs' motion is ballistic, the shape anisotropy is higher and  ranges between about 1.5 and 1.8.}
\label{fig:gyration}
\end{figure}

We next examine how collective SPP dynamics deform the enclosure by analyzing its shape anisotropy. 
The anisotropy is quantified by the ratio
$\langle\lambda_M/\lambda_m\rangle$, where $\lambda_M$ and $\lambda_m$ are, respectively, the larger and smaller eigenvalues of the gyration tensor,
\BE\label{eq:gyration_tensor}
G_{\alpha\beta}=\frac{1}{M}\sum_{i=1}^{M}\left(\alpha_i-\alpha_{CM}\right)\left(\beta_i-\beta_{CM}\right),
\EE
with $\alpha,\beta\in\{X,Y\}$ are the coordinates of the beads forming  the enclosure, and $(X_{CM},Y_{CM})$ are the coordinates of its center of mass. Figure~\ref{fig:gyration} shows that the dependence of $\langle\lambda_M/\lambda_m\rangle$ on $F_D$ follows the same overall trend as the diffusivity. In Regime I, the shape anisotropy is only slightly larger than unity, as expected for an approximately circular enclosure.
The shape anisotropy then increases rapidly over the range of $F_D$ where the Vicsek order parameter rises sharply, reaching a local maximum at $F_D\approx 16\,\varepsilon/r_c$.
This behavior is consistent with strong spatiotemporal heterogeneity in the collective motion of the SPPs, which gives rise to non--uniform and time--dependent stresses on the boundary. Such stresses induce anisotropic deformations of the enclosure and reduce the cancellation of forces on the enclosure, thereby enhancing net force fluctuations and the resulting diffusivity. As $F_D$ is increased,  reorganizations in the persistent vortical motion  occur less frequently, leading to stronger self--averaging of boundary stresses and, consequently, to  a reduction of both shape anisotropy and diffusivity.
As $F_D$ is further increased, the shape anisotropy then exhibits a second broad maximum in Regime III at $F_D\approx 32\,\varepsilon/r_c$, i.e., at about the same value of $F_D$, at which the diffusivity exhibits its second peak.

\begin{figure}[t]
  \begin{center}
	\includegraphics[scale=0.49]{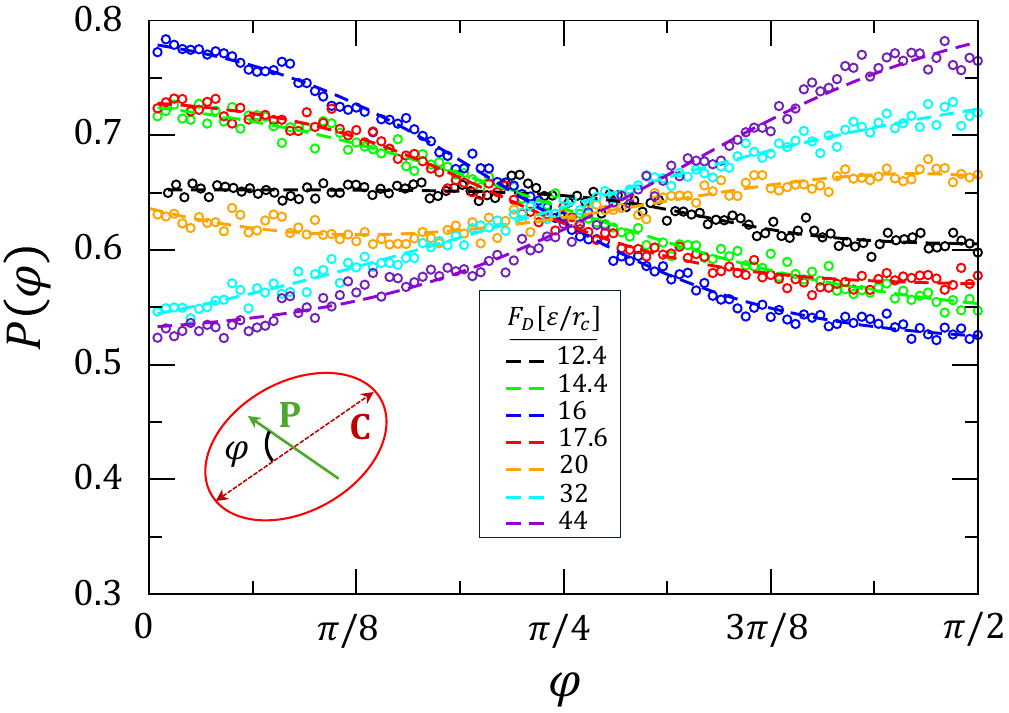}
   \end{center}
\caption{Normalized distribution of the acute angle $\varphi$ (shown schematically in the inset) between the net momentum of the SPPs, ${\mathbf P}$, and the long axis of the enclosure, ${\mathbf C}$, for different values of the driving force $F_D$. Results correspond to a flexible enclosure with $k_C=2000\,\varepsilon/r_c^2$.}
\label{fig:majoraxis_momentum}
\end{figure}

Having established that heterogeneity in the collective SPP dynamics leads to anisotropic deformation of the enclosure and enhanced diffusion, we next examine how the orientation of this deformation correlates with collective SPP motion.
This coupling is quantified by the distribution
$P(\varphi)$, where $\varphi$ is the acute angle between the net momentum of the SPPs, ${\mathbf P}$, and the enclosure’s long axis ${\mathbf C}$, defined as the eigenvector of the gyration tensor corresponding to the largest eigenvalue. Figure~\ref{fig:majoraxis_momentum} shows that $P(\varphi)$ is peaked at $\varphi=0$ in Regime II, indicating that the net SPP momentum is predominantly aligned with the enclosure’s long axis. This alignment is strongest at  $F_D\approx 16\,\varepsilon/r_c$, corresponding to the driving force at which the Vicsek order parameter reaches its maximum (see Fig.~\ref{fig:snapshots_order_param_flexible}).  As $F_D$ is increased further within Regime II, a second peak emerges at $\varphi=\pi/2$ (e.g., $F_D=20\,\varepsilon/r_c$).
At larger values of $F_D$, $P(\varphi)$ becomes unimodal again, now peaking at $\varphi=\pi/2$, indicating that the enclosure preferentially elongates perpendicular to the net SPP momentum.
This behavior is consistent with
a finite--size statistical effect: as vorticity reversals become less frequent, the larger number of SPPs distributed along the enclosure’s long axis leads to stronger 
self--averaging of momentum components in that direction, thereby biasing the net momentum toward the short axis.
This elongation perpendicular to the flock velocity may also be understood from a simple geometric argument.
The constraining boundary, being a passive ring, moves more slowly than the SPPs. A coherently moving flock of SPPs  that is constrained along its direction of motion must relax by spreading perpendicular to that direction. When this lateral spreading is arrested by an elastic boundary, it manifests as a steady ellipse whose long axis is perpendicular to the flock velocity.

\begin{figure}[t]
  \begin{center}
	\includegraphics[scale=0.67]{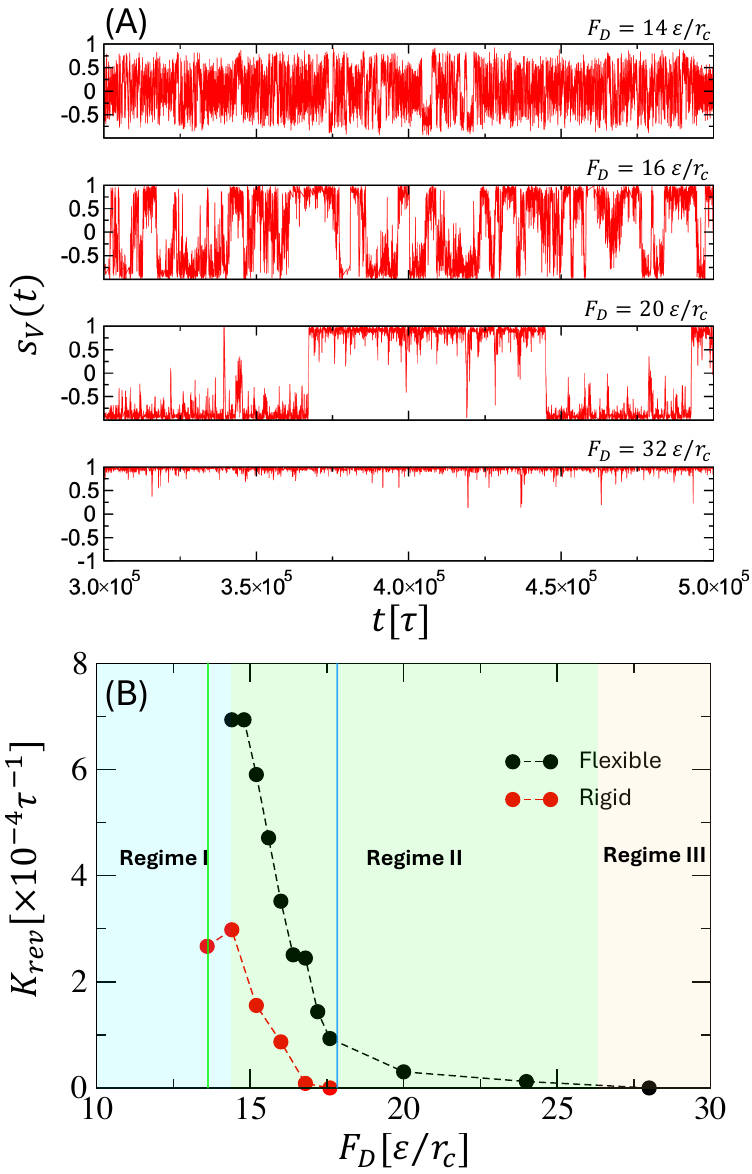}
   \end{center}
\caption{(A) Vortical order parameter $s_V$ vs time in the case of flexible enclosure with $k_C=2000\,\varepsilon/r_c^2$ at different values of $F_D$.
(B) Rate of vorticity reversals $K_{rev}$ vs. motility force for the case of flexible enclosure with same $k_C$ as in (A) (black points) and for the case rigid enclosure (red points). The shaded regions correspond to Regime I and II in the case of flexible confinement. 
The vertical green (blue) line separates Regime I (II) from Regime II (III) in the rigid-confinement case.}
\label{fig:reversal_rate}
\end{figure}

We now focus on vorticity reversals in Regime II. Reversals in the direction of the vortical flow are characterized using a time-dependent vortical order parameter, $s_V(t)$, computed from SPPs near the boundary and defined as
\BE \label{eq:time_dependent_vortical_order_parameter}
s_V(t)=\frac{1}{{\cal P}_{b}(t)}\sum_{l \in {\cal B}(t)}{\sigma_{l}(t)},
\EE
where $\sigma_l(t)$ is defined in the same manner as in 
Eq.~(\ref{eq:vortical-order}), ${\cal B}(t)$ denotes the set of SPPs whose centers of mass at time $t$ lie within an annular region of width
$15\,r_c$ near the confining boundary, and ${\cal P}_{b}(t)$ is the number of SPPs in ${\cal B}(t)$.
Figure~\ref{fig:reversal_rate} shows that the rate of reversals, $K_{rev}$, decreases with $F_D$ for both rigid and flexible confinements. However, $K_{rev}$ is substantially lower in the case of rigid confinement, with the enclosure area matched to that of the flexible case. This indicates that boundary flexibility strongly promotes vorticity reversals.

\begin{figure}[t]
  \begin{center}
	\includegraphics[scale=0.6]{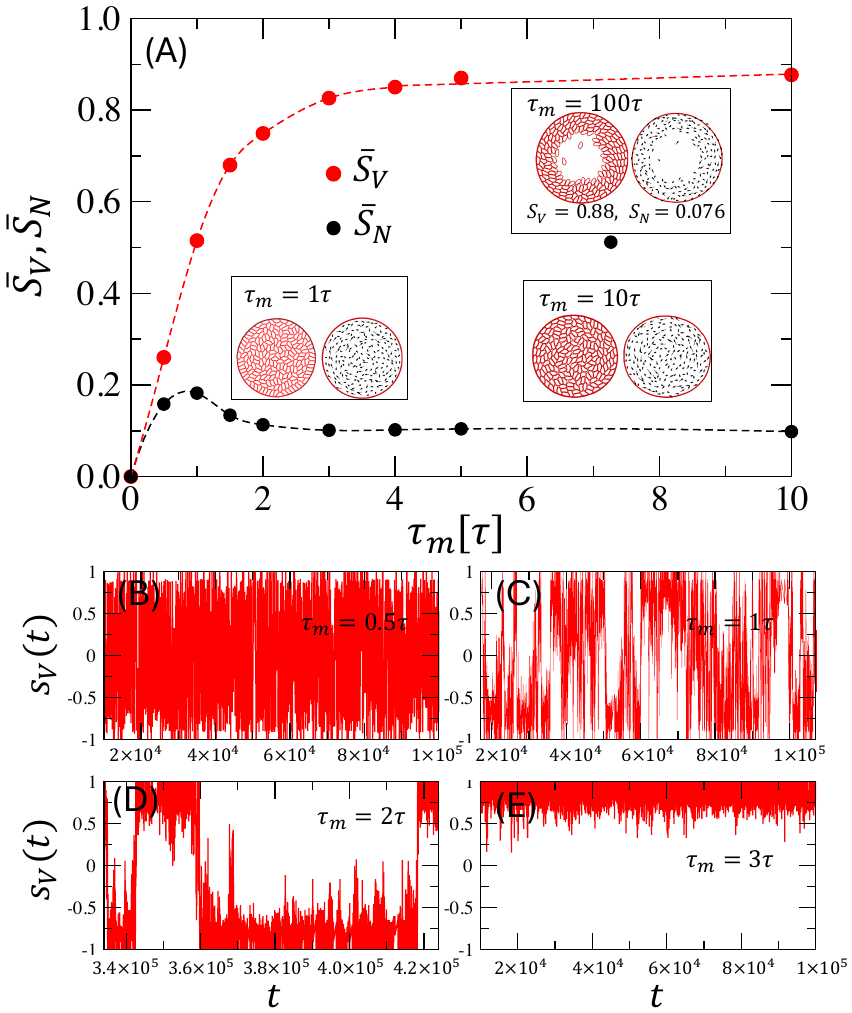}
   \end{center}
\caption{(A) Vortical and Vicsek order parameters vs. $\tau_m$. (B--E) Time dependence of the vortical order parameter $s_V(t)$ for a layer of SPPs near the boundary at values of $\tau_m$ ranging between 0.5 and $3\,\tau$. Data in the figure are for $F_D=16\,\varepsilon/r_c$ in the case of flexible confinement with $k_C=2000\,\varepsilon/r_c^2$.}
\label{fig:tau_m}
\end{figure}

\begin{figure}[t]
  \begin{center}
	\includegraphics[scale=0.5]{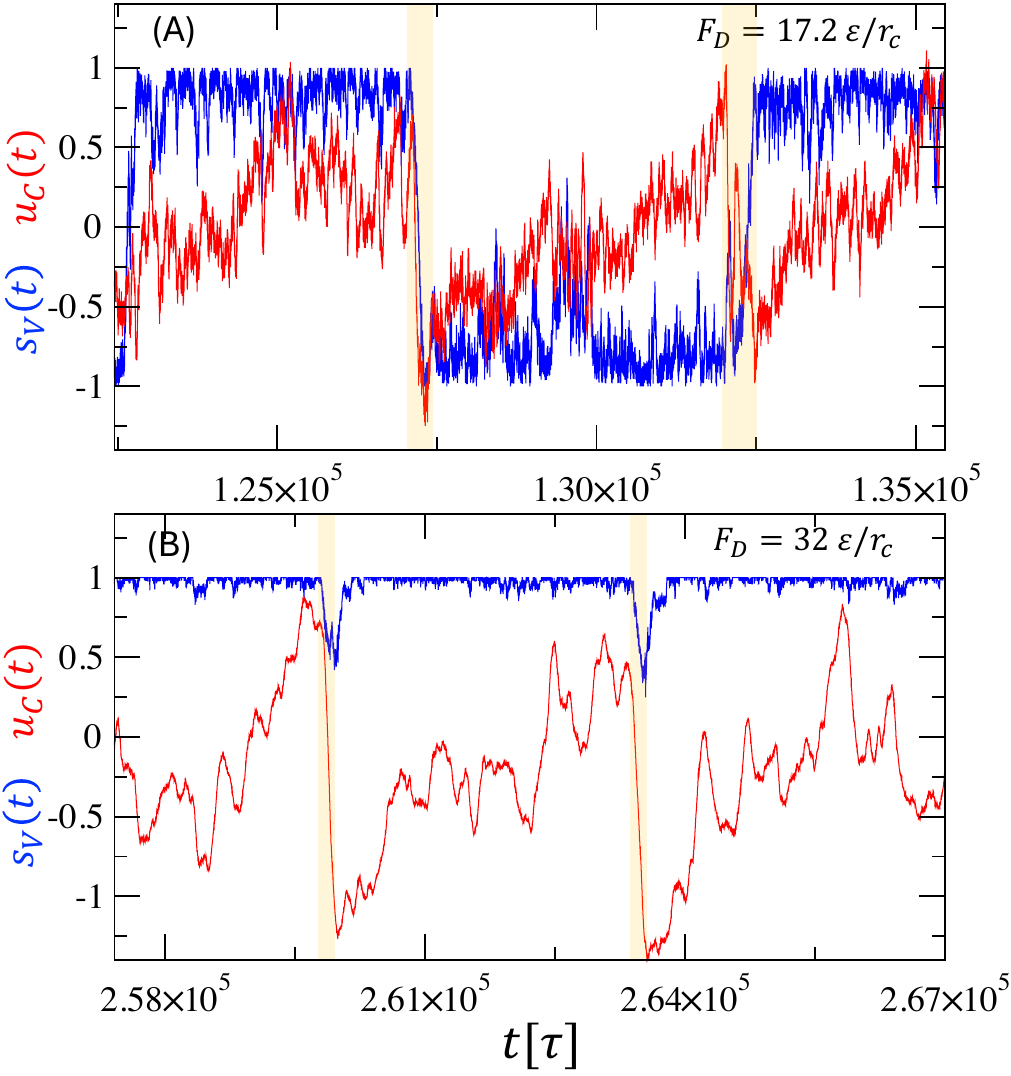}
   \end{center}
\caption{ 
Time evolution of the vortical order parameter $s_V(t)$ (blue) and the  boundary's elastic energy (red) in the case where $k_C=2000\,\varepsilon/r_c^2$. 
The elastic energy is shown in rescaled form,
${u}_C(t)=\alpha\left({\cal U}_C(t)/\bar{\cal U}_C-1\right)$, where $\bar{\cal U}_C$ is the steady-state time-averaged elastic energy of the boundary. (A) corresponds to $F_D=17.2\,\varepsilon/r_c$,  with $\alpha=8$ and $\bar{\cal U}_C=23\,509\,\varepsilon$. (B) corresponds to  $F_D=32\,\varepsilon/r_c$, with $\alpha=2.4$ and $\bar{\cal U}_C=61\,959\,\varepsilon$. The rescaling  is used solely so to visualize both quantities on the same scale. Only a representative time window is shown. 
The orange-shaded regions highlight episodes in which stress release is accompanied by a loss of vortical coherence. In Regime II ($F_D=17.2\,\varepsilon/r_c$), such events frequently culminate in full reversals of the vorticity, whereas in Regime III ($F_D=32\,\varepsilon/r_c$), stress release leads only to transient vortical incoherence without successful reversals. }
\label{fig:vortical_stress_timeseries}
\end{figure}

To determine the origin of vorticity reversals, we performed a series of simulations for flexible confinement in which the persistence time $\tau_m$ in Eq.~(\ref{eq:historical_velocity}) was systematically varied between 0 and $100\,\tau$ at fixed driving force $F_D=16\,\varepsilon/r_c$. Figure~\ref{fig:tau_m} shows that the vortical order parameter increases monotonically with $\tau_m$, while the reversal rate decreases as $\tau_m$ increases. This behavior reflects the fact that larger $\tau_m$ enhances the persistence of unidirectional motion of the SPPs, thereby suppressing spontaneous changes in individual particle directions. Qualitatively similar trends are observed for rigid confinement. 
In the limiting case where the SPPs lose the ability to reverse their direction of motion ($\tau_m \rightarrow \infty$), the system no longer supports reversal-mediated reorganization. Instead, a strongly polarized state emerges in which SPPs segregate into two dominant groups localized near opposite sides of the enclosure and moving in opposite directions. The imbalance between these counter-propagating clusters generates a net force on the boundary, resulting in sustained ballistic drift of the entire enclosure (Movie 6).
Moreover, when thermal noise is switched off, vorticity reversals are no longer observed. Under soft confinement, the system exhibits ballistic motion (Movie 7), whereas under rigid confinement it settles into a persistent unidirectional vortical state (Movie 8). These results indicate that vorticity reversals originate from noise-enabled fluctuations intrinsic to the particle dynamics. Boundary flexibility amplifies the likelihood of reversal events through stress–boundary feedback, but is not sufficient to generate reversals in the absence of noise.

We next examine the coupling between the vortical motion of the SPPs and boundary deformation by analyzing the time-resolved evolution of the vortical order parameter and the elastic energy of the boundary (see Fig.~\ref{fig:vortical_stress_timeseries}). In Regime II, shown for $F_D=17.2$ and $32\,\varepsilon/r_c$ in Fig.~\ref{fig:vortical_stress_timeseries}(A), the elastic energy of the boundary typically increases gradually during intervals of sustained vortical motion. This reflects the buildup of stress exerted on the boundary by the circulating SPPs. In contrast, rapid drops in the elastic energy coincide with a loss of vortical coherence, as indicated by a decrease in the absolute value of $s_V$. As shown in Fig.~\ref{fig:vortical_stress_timeseries}(A), such stress-release events frequently develop into complete reversals of the vorticity in Regime II.

In the case of rigid confinement, transient reductions in vortical order are also observed. However, they rarely develop into full reversals, suggesting that, in this case, the two oppositely rotating states are separated by a relatively large effective energy barrier. For a flexible boundary, stress-release events are accompanied by localized deformations that enhance boundary–SPP interactions and increase velocity fluctuations. These fluctuations can temporarily lower the barrier between the two vortical states, allowing partial loss of coherence to grow into a complete reversal. At high driving forces ($F_D \gtrsim 26\,\varepsilon/r_c$, Regime III), the vortical motion becomes highly persistent and the barrier between opposite circulation states increases. Consequently, although stress-release events still occur, they no longer lead to reversals (Fig.~\ref{fig:vortical_stress_timeseries}(B)). This indicates that elastic stress relaxation does not by itself cause vorticity switching, but instead facilitates it by transiently lowering the barrier against reversal in Regime II.

\begin{figure}[t]
  \begin{center}
	\includegraphics[scale=0.675]{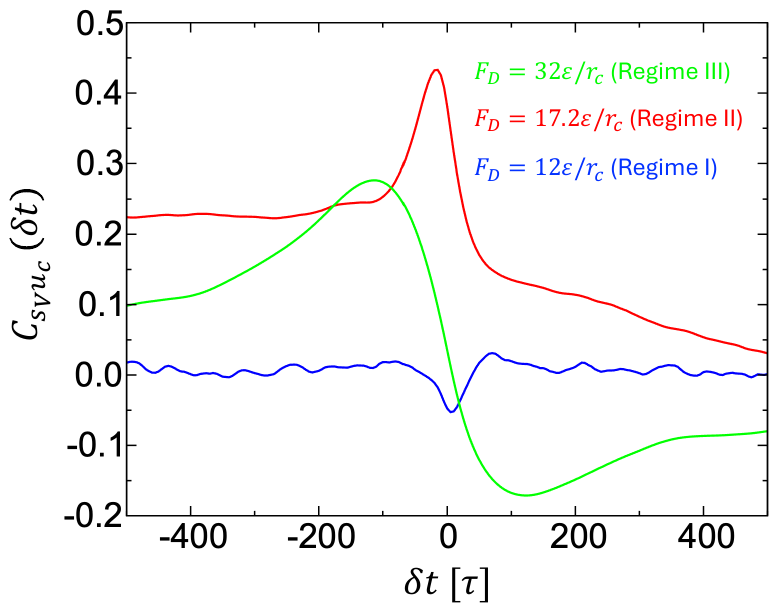}
   \end{center}
\caption{\cbb Normalized cross-correlation function $C_{s_V u_c}(\delta t)$ between 
the boundary elastic energy and the absolute value of the vortical order parameter, 
defined in Eq.~(\ref{eq:cros_correl}), for representative values of $F_D$ in 
Regimes~I, II, and III, in the case of a flexible enclosure with 
$k_C = 2000\,\varepsilon/r_c^2$.}
\label{fig:cross-correl}
\end{figure}

{\cbb
To directly quantify the causal relationship between boundary elastic energy and vortical coherence suggested by Fig.~\ref{fig:vortical_stress_timeseries}, we computed the normalized cross-correlation function
\begin{equation}
C_{s_V u_c}(\delta t) = \frac{\langle \delta {\cal U}_c(t)\, \delta S_V(t+\delta t)\rangle_t}{\sqrt{\langle \delta {\cal U}_c(t)^2\rangle_t \langle \delta S_V(t)^2\rangle_t}},
\label{eq:cros_correl}
\end{equation}
where $\delta {\cal U}_c(t) = {\cal U}_c(t) -  {\bar {\cal U}}_c$ and $\delta S_V(t) = S_V(t) - {\bar S}_V$. Results for representative values of $F_D$ in Regimes~I, II, and III are shown in Fig.~\ref{fig:cross-correl}. In Regime~I ($F_D = 12\varepsilon/r_c$), $C_{s_V u_c}(\delta t)$ is essentially zero for all $\delta t$, as expected in the absence of coherent motion. The picture changes markedly in Regime~II ($F_D = 17.2\varepsilon/r_c$). Here, $C_{s_V u_c}(\delta t)$ is positive and slowly rising for large negative $\delta t$, reflecting the gradual buildup of elastic energy during intervals of sustained vortical motion, over which $S_V(t)$ remains roughly constant. The function then rises sharply to a peak at $\delta t \approx -20\tau$: the negative lag indicates that the drop in boundary elastic energy precedes the loss of vortical coherence by approximately $20\tau$, establishing a clear causal ordering between stress release and vortical disruption. For positive $\delta t$, $C_{s_V u_c}$ decays rapidly and turns negative, since a drop in elastic energy below its mean is followed, after the reversal, by recovery of $S_V(t)$ to above its mean as the SPPs re-establish coherent circulation. In Regime~III ($F_D = 32\varepsilon/r_c$), the picture is qualitatively different. The peak is broader and shifted to $\delta t \approx -200\tau$, and the negative excursion is more pronounced. In this regime, stress-release events do produce transient dips in $S_V$, as seen in Fig.~\ref{fig:vortical_stress_timeseries}(B), and the cross-correlation confirms that these dips are systematically preceded by a drop in elastic energy. However, as discussed above in relation to Fig.~\ref{fig:vortical_stress_timeseries}, the vortical motion in Regime~III is sufficiently coherent that stress-release events are unable to fully reverse the circulation direction: $S_V$ dips only transiently before recovering to above its mean, giving rise to the negative lobe of $C_{s_V u_c}$ at positive $\delta t$. The cross-correlation thus puts the stress--boundary feedback on a quantitative footing: elastic stress builds up during coherent vortical motion, its release triggers a transient loss of vortical coherence, and coherence is subsequently re-established. Whether this process culminates in a full reversal of the circulation direction, as in Regime~II, or merely a transient disruption, as in Regime~III, depends on the degree of vortical coherence and the height of the effective barrier between the two circulation states. The precise microscopic mechanism by which stress release destabilizes the vortical flow, however, remains an open question.
}

The sensitivity of the collective dynamics to boundary flexibility is further examined by varying the elasticity stiffness $k_C$ of the enclosure. Figure~{\cbb S5} in the ESI shows the time-averaged Vicsek and vortical order parameters as functions of $k_C$ at representative driving forces in Regimes II and III. At $F_D = 16\,\varepsilon/r_c$ (Regime II), increasing $k_C$ up to about $2000\varepsilon/r_c^2$ enhances both $\bar S_N$ and $\bar S_V$, indicating that moderate increases in boundary stiffness promote more coherent collective motion. With further increase of $k_C$, $\bar S_N$ remains nearly constant, while $\bar S_V$ increases only weakly. We also found the rate of vorticity reversals decreases with $k_C$. Figure {\cbb S5} shows that at $F_D = 40\,\varepsilon/r_c$ (Regime III), decreasing $k_C$ (i.e., increasing boundary flexibility) drives a qualitative change in SPPs collective dynamics from unidirectional vortical motion to ballistic motion, indicating that boundary deformability lowers the activity transition from Regime III to Regime IV. These results demonstrate that boundary flexibility acts as a continuous control parameter that modulates the balance between vortical and polar collective dynamics, rather than selecting a single finely tuned state.
{\cbb In the present work, we focused on the role of the stretching elasticity $k_C$, which primarily controls the large-scale deformability and expansion of the enclosure under active stresses. Variations of the bending modulus $\kappa_C$ are expected to mainly affect local curvature fluctuations and quantitatively shift the dynamical boundaries without changing the qualitative feedback mechanisms discussed here.}

The emergence of polar, ballistic motion at high driving forces can be rationalized in terms of how the stresses exerted by the SPPs are distributed on the boundary during vortical motion. In the vortical state, circulating SPPs generate a persistent outward normal stress on the enclosure that increases with $F_D$, leading to a gradual expansion of the enclosure area (see Fig.~{\cbb S3} in the ESI). This outward stress is balanced by the elastic response of the enclosure. 
Sustaining such a vortical state beyond a threshold value of $F_D$ therefore requires the boundary to balance increasingly large, spatially distributed stresses through its elastic response.

In the polar state, by contrast, SPPs accumulate predominantly on one side of the enclosure, which strongly reduces the normal stress on its opposite side. 
At the same time, global alignment   of SPP velocities 
redirects a substantial fraction of the active forcing into directed motion of the
enclosure rather than its elastic deformation. Active stresses are therefore no longer
distributed isotropically along the boundary but are instead organized into a net propulsive force. 
Consistent with this redistribution of stresses, the enclosure area decreases upon entering Regime~IV, as demonstrated by Fig.~{\cbb S3} in the ESI.

\begin{figure}[t]
  \begin{center}
	\includegraphics[scale=0.51]{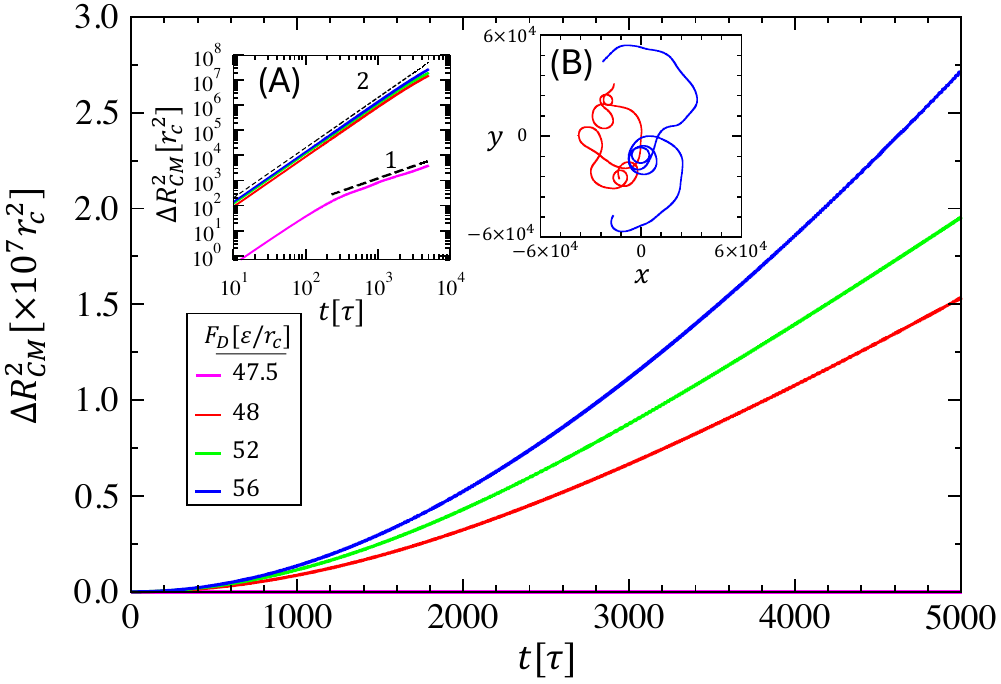}
   \end{center}
\caption{MSD of the enclosure's center of mass vs time in Regime III (magenta curve) and Regime IV (red, green, and blue curves) in the case where $k_C=2000\,\varepsilon/r_c^2$. Inset (A): MSD vs time in a double logarithmic plot. The slope of 1 of the magenta line indicates diffusive motion of the enclosure in Regime III. The slope of 2 in Regime IV indicates a ballistic propulsion of the enclosure. Inset (B): Trajectories of the enclosure's center of mass at $F_D=48\varepsilon/r_c$ (red curve) and $F_D=56\varepsilon/r_c$ (blue curve).
}
\label{fig:regime4_propulsion}
\end{figure}

As shown by the snapshots in Fig.~\ref{fig:snapshots_order_param_flexible} and by Movie 3 in the ESI, the enclosure in Regime~IV elongates predominantly in a direction perpendicular to the net momentum of the SPPs. A microscopic explanation for this behavior follows from examining how SPPs reorganize near the leading edge in the polar state. If the enclosure elongates primarily along the direction of the net SPPs momentum, only a limited number of SPPs would be in direct contact with the boundary at the front. Because SPPs are flexible and fairly compressible, particles in successive smectic-like rows behind the leading edge can squeeze forward and join the contact layer. This increases the number of SPPs that directly push on the boundary and widens the region of contact at the front, causing the leading edge to expand laterally while elongating the enclosure perpendicular to the direction of motion.

We finally examine the directed motion of the enclosure induced by the collective polar alignment of the SPPs in Regime IV. 
The MSD of the enclosure’s center of mass, shown in Fig.~\ref{fig:regime4_propulsion}, grows quadratically with time, demonstrating ballistic transport of the enclosure. The prefactor of the quadratic scaling increases with  
$F_D$, indicating that the enclosure's propulsion speed rises with activity. 
Although the motion of the enclosure is ballistic, the propulsion direction evolves slowly with time, resulting in curved trajectories, as shown by two representative trajectories in Fig.~\ref{fig:regime4_propulsion}(B).
This gradual reorientation originates from fluctuations in the spatial distribution of SPPs near the leading edge. Because the enclosure deforms and elongates perpendicular to the instantaneous direction of motion, even small {asymmetries} 
in the number of SPPs {exerting forces on either side of the leading edge} 
generate a weak, time-dependent torque that continuously redirects the propulsion axis.
This coexistence of
rapid ballistic translation 
and slower directional reorientation is a defining feature of Regime~IV.

\section{Conclusion}

In this work, we investigated the collective dynamics of {reversal-capable, environment-sensitive} 
SPPs confined within deformable enclosures, with particular emphasis on the interplay between boundary flexibility and the organization of active motion,
using coarse-grained molecular dynamics simulations in two dimensions. We identified a sequence of distinct dynamical regimes as the motility force is increased, ranging from thermally dominated motion to vortical flow with stochastic reversals, followed by persistent vortical motion and, at still higher activity, persistent polar motion of the SPPs accompanied by ballistic propulsion of the enclosure.

In the vortical regime, circulating SPPs exert a spatially distributed outward normal stress on the boundary, which is balanced by elastic deformation of the enclosure. Fluctuations in the spatial organization and collective motion of the SPPs lead to non-uniform boundary stresses, giving rise to anisotropic enclosure shapes and enhanced diffusive propulsion. As either the driving force $F_D$ or the persistence time $\tau_m$ increases, the relative influence of thermal fluctuations diminishes, reversals are suppressed, and both the diffusivity and shape anisotropy exhibit non-monotonic behavior reflecting changes in stress heterogeneity and persistence.

At sufficiently high activity, the system undergoes a {\cbb discontinuous} transition to a polar, ballistic regime characterized by global alignment of SPP velocities and sustained propulsion of the enclosure. In this regime, SPPs accumulate at the leading edge, redistributing active stresses and 
changing the nature of the boundary stresses from tension dominated to bending dominated. The enclosure elongates perpendicular to the direction of motion due to local reorganization within the contact layer, where flexible SPPs squeeze into the leading edge and increase the contact width. The resulting propulsion is ballistic on accessible timescales, with a slow rotational drift arising from fluctuations in the asymmetric distribution of SPPs at the front.

It is important to note that the present study is performed in two dimensions and employs a specific implementation of reversal-capable SPPs modeled as deformable ring polymers with nematic polarity and finite velocity memory. 
In addition, {\cbb the model describes elastic rings confining discrete active particles, in the spirit of recent experiments on synthetic cell-mimics confined within deformable boundaries~\cite{arora2024}, and does not include explicit hydrodynamic interactions and incompressibility constraints. It should therefore not be interpreted as a model for active fluid droplets~\cite{tiribocchi2023a,tiribocchi2023b,tiribocchi2023c}, in which the deformable boundary is the fluid-fluid interface of an incompressible droplet, rather than a distinct elastic structure enclosing discrete active particles.}

{\cbb While quantitative details of phase boundaries, vortex structure, cluster-size measures, and deformation amplitudes may depend on system size, particle geometry, the number of beads used to discretize the SPPs, propulsion rules, and boundary constraints, the central mechanisms identified here are expected to be more general, as we now summarize.} 
First, boundary deformability strongly amplifies vorticity reversals by coupling stress-release events to fluctuations in collective motion. Second, deformability reorganizes active stresses, enabling a transition from spatially distributed elastic loading in the vortical state to ballistic motion of the enclosure. Third, the environmental sensitivity of the SPPs, encoded through their ability to reverse direction and their finite persistence time $\tau_m$, provides an internal dynamical channel that modulates the effective barrier between symmetry-related vortical states. By tuning the balance between memory-based reorientation and persistent motion, this sensitivity regulates whether stress fluctuations trigger reversals or instead stabilize coherent circulation.

These mechanisms rely on stress-boundary-memory feedback rather than on the specific microscopic details, and should therefore persist in other classes of interacting SPPs under soft confinement.
{\cbb Although reversal-capable bacteria such as \textit{Vibrio alginolyticus} and \textit{Myxococcus xanthus} are well-established biological realizations of non-tumbling swimmers, to our knowledge no experimental studies have yet examined their collective behavior under soft confinement. In our model, direction changes are not governed by a fixed stochastic clock but emerge from the recent swimming history of each SPP as shaped by confinement and inter-particle interactions, making the reversal rate inherently sensitive to the local environment. A related biological example is provided by smooth-swimming \textit{E. coli} mutants lacking CheY ($\Delta$\textit{cheY}), which cannot switch flagellar motor direction yet exhibit frequent reversals in collective swarms through a purely mechanical mechanism that requires no motor switching~\cite{Wu2020}. Engineered non-tumbling \textit{E. coli} confined within elastic vesicles would therefore provide a promising experimental platform for probing the memory-based reversal dynamics identified here, with the effective persistence time controllable through cell density, confinement geometry, or surface conditions.}

Experimental realizations of active particles confined within deformable boundaries or elastic droplets would provide a promising platform to probe these stress-mediated mechanisms and to explore how boundary mechanics can be used to regulate collective behavior and emergent boundary propulsion. In particular, environmental sensitivity could be probed by tuning particle persistence times, reversal capability, or activity fluctuations via external fields, chemical cues, or light-activated control, thereby experimentally accessing the stress-boundary-memory feedback mechanism highlighted in this work.

\section{Acknowledgments}

All simulations were performed on the University of Memphis BigBlue high-performance computing cluster. PBSK acknowledges the financial support provided by Anusandhan National Research Foundation (ANRF) under J. C. Bose grant (ANRF/JBG/2025/000187/PS). ML gratefully acknowledges the hospitality of the Indian Institute of Technology Madras, where part of this work was carried out.

\bibliographystyle{rsc}
\bibliography{citations}

\end{document}